\documentclass[a4paper,aps,prl,twocolumn,floatfix,showpacs,groupedaddress,linenumbers]{revtex4}
\usepackage{graphics}
\usepackage{epsfig}
\usepackage{times}
\usepackage{xcolor}   
\usepackage{amsfonts}
\usepackage{amsmath}
\usepackage{amssymb}
\usepackage{amsthm}
\usepackage{hyperref} 
\usepackage{wasysym}
\usepackage{bm} 
\usepackage{subfig} 
\newcommand{\sgn}{\mathop{\mathrm{sgn}}}

\def\dfrac#1#2{{\displaystyle\frac{#1}{#2}}}

\def\lsim{\mathrel{\rlap{\lower4pt\hbox{\hskip1pt$\sim$}}
    \raise1pt\hbox{$<$}}}         
\def\gsim{\mathrel{\rlap{\lower4pt\hbox{\hskip1pt$\sim$}}
    \raise1pt\hbox{$>$}}}         

\newcommand{\beq}{\begin{equation}}
\newcommand{\eeq}{\end{equation}}
\newcommand{\bea}{\begin{eqnarray}}
\newcommand{\eea}{\end{eqnarray}}

\newcommand{\ie}{{\it i.\,e.\,\,}}

\def\lsim{\:\raisebox{-0.5ex}{$\stackrel{\textstyle<}{\sim}$}\:}
\def\gsim{\:\raisebox{-0.5ex}{$\stackrel{\textstyle>}{\sim}$}\:}

\newcommand\sch{Schr$\ddot{\rm o}$dinger~}

\newcommand\bloch{Bloch~}

\begin{document}
\textheight=23.8cm
\title{\Large Topological p-n junctions in helical edge states}
\author{Disha Wadhawan$^1$, Poonam Mehta$^2$ and Sourin Das$^{1,3}$}
\email{dwadhawan@physics.du.ac.in, pm@jnu.ac.in, sdas@physics.du.ac.in}
\affiliation{$^1$ Department of Physics and Astrophysics, University of Delhi, Delhi 110007\\
$^2$ School of Physical Sciences, Jawaharlal Nehru University, New Delhi 110067 \\
$^3$ Max-Planck Institute for the Physics of Complex Systems, 01187 Dresden.}
\date{\today}

\renewcommand{\thefootnote}{\fnsymbol{footnote}}
\setcounter{footnote}{-1}
%
\begin{abstract}
 Quantum spin Hall effect is endowed with topologically protected edge modes with gapless Dirac spectrum.  Applying a magnetic field locally along the edge leads to a gapped edge spectrum with opposite parity for winding of spin texture for conduction and valence band. Using Pancharatnam's prescription for geometric phase it is shown that mismatch of this parity across a $p$-$n$ junction, which could be engineered into the edge by electrical gate induced doping, leads to a phase dependence in the two-terminal conductance which is purely topological (0 or $\pi$). This fact results in a ${\mathbb{Z}}_2$ classification of such junctions with an associated duality. Current asymmetry measurements which are shown to be robust against electron-electron interactions are proposed to infer this topology. 
\end{abstract}
%
\maketitle
\begin{figure}[htb!]
 \includegraphics[width=1.0\linewidth]{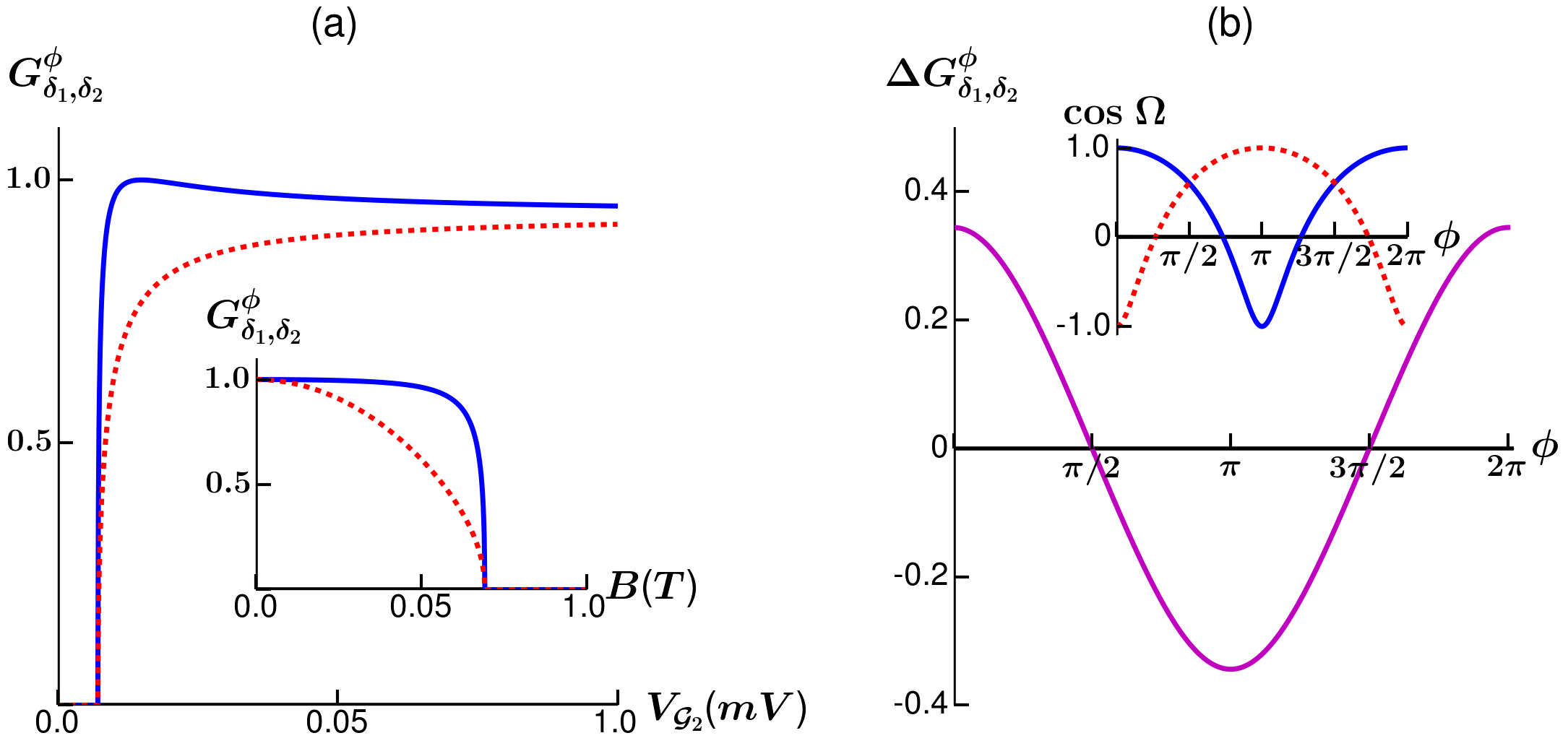}
 \caption{(a)Landauer Conductance $G^{\phi}_{\delta_1,\delta_2}$ in units $e^2/h$ is plotted as a function of chemical potential, $V_{{{\cal G}}_2}$ (or $\delta_{2}$) in case of $n$-$n$ (blue curve) and $n$-$p$ (red curve) junctions for $\delta_{1}=0.15 mV$ (fixed) , $\phi=0$  with $B=50mT$ and the inset depicts variation of $ G ^{\phi}_{\delta_1,\delta_2}$ in units $e^2/h$ with applied magnetic field, when, $V_{{{\cal G}}_2}=0.1 mV$ (b)shows the variation in the difference of conductance ($\Delta {\cal G}^{\phi}_{\delta_1,\delta_2}$) with the rotation of magnetic field ($\phi$)and the inset shows variation of cosine of $\Omega$ in case of the two junctions with $\phi$ .} 
 \label{fig1}
\end{figure}

{{\sl{\underline{Introduction} :}}} In a seminal paper\cite{Berry:1984jv}, Berry introduced the cyclic and adiabatic geometric phase that had implications across disciplines\cite{shaperebook} ranging from high energy physics to condensed matter physics. It was soon realized that both the conditions of adiabaticity\cite{nakagawa,Aharonov:1987gg} as well as  cyclicity of evolution\cite{sam} were not at all necessary and the geometric phase was a property of the Hilbert space itself. It turned out that a generalized version of this idea was anticipated\cite{ramaseshan1986,Berry1987} by Pancharatnam in his pioneering work on interference of classical light in distinct states of polarization\cite{Pancharatnam:1956}. Pancharatnam's connection (or rule) gives a natural way to compare the relative phases between any two non-orthogonal states, $|{\cal A}\rangle$ and $|{\cal B}\rangle$. If $\langle {\cal A} | {\cal B} \rangle$ is real and positive, they are said to be ``in phase'' or ``parallel''. For a two-state quantum system, if we consider three arbitrary non-orthogonal states $|{\cal A}\rangle,|{\cal B}\rangle,|{\cal C} \rangle$ such that pair-wise $|{\cal A}\rangle,|{\cal B}\rangle$ and $|{\cal B}\rangle,|{\cal C}\rangle$ are in phase, it turns out $|{\cal C}\rangle$ is not necessarily in phase with $|{\cal A}\rangle$. Deviation of $|{\cal C}\rangle$ being in phase with $|{\cal A}\rangle$ is quantified in terms of phase of the complex number $\langle {\cal A} |{\cal B} \rangle \langle {\cal B}|{\cal C}\rangle \langle {\cal C}|{\cal A}\rangle$. This phase is given by half the solid angle of the geodesic triangle ${\cal A} {\cal B} {\cal C}$ subtended at the centre of the \bloch sphere.\\
We show that transport in helical 1D electron gas  (1DEG)\cite{Wu2006} with Dirac-like spectrum is dominated by geometric phase of the type described above when a gap is opened up in the spectrum due to application of magnetic field while transport is induced in the gapped spectrum by pure electrical doping. The spin-momentum locked helical states are believed to be of potential importance for future spintronics device applications\cite{yoko2014}, hence an understanding of transport from a geometric phase point of view in presence of standard probes such as magnetic field or electric field could provide useful guidelines for efficient manipulation of these states for such applications. In this article we show that electrical transport across $p$-$n$, $n$-$p$, $p$-$p$, $n$-$n$ junctions designed into the helical edge have a  ${{\mathbb{Z}}}_2$ topological classification which essentially stems from Pancharatnam's prescription of geometric phase and has its origin in spin-momentum locked nature of the gapless spectrum. Here p and n corresponds to hole and electron type doping. We also show that this topological description of the junction facilitates an effective conversion of $p$-$n$  $\leftrightarrow$ $p$-$p$ (as far as the electrical transport properties of these junctions are concerned) not by changing the doping but by suitably manipulating the magnetic field acting on them, hence opening up new possibilities for manipulation of these junctions. Finally we discuss the experimental feasibility of our proposal when the 1D helical state is hosted on the edge of a quantum spin Hall state that was experimentally observed in HgTe/CdTe quantum wells\cite{Konig2007}.\\
{{\sl{\underline{Model} :}}} 
 We consider a 1-D helical state physically lying along the $x$-axis with its spin-orbit (SO) field pointing along  $z$-axis which is exposed to external applied magnetic field and gate electrodes imposed electric field. The Hamiltonian  is given by
 \begin{eqnarray}
{\cal H} = - i \hbar v_{F} \sigma_{z}\partial_{x} + g\mu_{B}\left(\vec{S} \cdot \vec{B}(x,y)\right) 
    + eV_{{{\cal G}}}(x)~,
\label{H}
 \end{eqnarray}
where $\hbar = {h}/{2\pi}$ is the reduced Planck constant, $v_{F}$ is the Fermi velocity, $\sigma_{x,y,z}$ are the Pauli matrices, $\vec{S}=\frac{\hbar}{2}\vec{\sigma}$ is the spin operator, $g$ is the Land\'e-g factor for the electron, $\mu_{B}$ is the Bohr magneton, $e$ is the electronic charge and $V_{{{\cal G}}}(x)$ is the  applied gate voltage. Gating induced electrostatic potentials  $V_{{{\cal G}}}=V_{{{\cal G}}_1},V_{{{\cal G}}_2}$ are assumed to be acting independently on the two halves  ($x \in [-\infty,0]$ and $x \in [0,\infty]$) respectively. 
These electrostatic potentials allow for independent tuning of each patch either to electron-type or to hole-type doping. The external applied magnetic field profile acting on the 1-D helical state is taken to be 
\begin{equation}
\vec{B} (x,y) =
\left\{ \begin{array}{rcl}
B \hat{x} & \mbox{for}
& -\infty \leq x < 0 \\ 
(B\cos\phi)\hat{x}+(B\sin\phi)\hat{y}  & \mbox{for} & 0 < x \leq \infty\,.
\end{array}
\right.
 \label{Bfield}
\end{equation} 
Application of magnetic field along any direction in $x$-$y$ plane opens up a gap in the dispersion\cite{Soori2012} while selecting different in-plane directions for the $\vec{B}$-field in the two patches allows for manipulation of the Pancharatnam geometric phase in a desired way as we will see later.\\
{{\sl{\underline{Landauer conductance} :}}}  The energy spectrum for the problem in each semi-infinite patch is given by  $E_{k}= e V^{}_{\cal{G}}\pm v_{F}^{} \sqrt{k^{2}+b^2}$ where $b=g\mu_{B}B/ 2v_{F}$ and $k$ is the momentum of the electron. The corresponding momentum dependent eigen-spinor is given by 
\begin{equation}
\psi_{k,\epsilon} = (\epsilon / N_{\epsilon}^{}) 
\left[ \begin{array}{cc} ( k_{}^{} +  \sqrt{k_{}^{2} + b^{2}_{}})/(b e^{i \phi}_{})   & 1  \end{array}  \right]^{T}~. 
\label{wf}
\end{equation} 
where $N_{\epsilon}$ is the normalization constant, $\epsilon=\pm$ stands for conduction and valence band respectively. Now by demanding continuity of the plane-wave solution of the \sch equation at the junction we obtain the Landauer conductance in the linear response limit which is expressed  in terms of transmission probability $(T^{\phi}_{V_{{{\cal G}}_1},V_{{{\cal G}}_2}})$ as 
\bea
{\cal G}^{\phi}_{V_{{{\cal G}}_1},V_{{{\cal G}}_2}}=\dfrac{e^2}{h}~ T^{\phi}_{V_{{{\cal G}}_1},V_{{{\cal G}}_2}}=\dfrac{e^2}{h} \left(\dfrac{\rho_{i}}{\rho_{t}}\right)  {\cal S} (\alpha,\beta,\gamma)~,
 \label{T}
\eea
where, ${\cal S}({\alpha,\beta,\gamma})=(1-\vert\alpha\vert^2)^2/(\vert\beta\vert^2+\vert\alpha\vert^2\vert\gamma\vert^2-2\vert\alpha\beta\gamma\vert\cos{\Omega})$, the  parameters $\alpha$, $\beta$  and  $\gamma$ are the spinor overlaps given by $\alpha=\langle \psi_{{r}}\vert\psi_{{i}}\rangle$, $\beta=\langle \psi_{{i}}\vert\psi_{{t}}\rangle$, $\gamma=\langle \psi_{{t}}\vert\psi_{{r}}\rangle$ where the indices $i,r,t$ stand for the  incident, reflected  and transmitted spinors respectively evaluated at the average Fermi energy across the junction. The exact expressions for  $\alpha,\beta$ and $\gamma$ would depend upon $V_{{{\cal G}}_1},V_{{{\cal G}}_2}$ which decides the doping on the two sides of the junction and the relative angle between the applied magnetic fields, $\phi$ via Eq.{\ref{wf}}. $\rho_{i}$ and $\rho_{t}$ are the density of states of the incident(reflected) branch and the transmitted branch given by $\rho_{i}^{}=\sqrt{k_{i}^{2}+b^2_{}}/ ( 2\, \pi\, v_{f}^{} k^{}_{i})$,  $\rho_{t}^{}=\sqrt{k_{t}^{2}+b_{}^2}/( 2\, \pi\, v^{}_{f} k^{}_{t})$.  And $\Omega$ is the Pancharatnam geometric phase as defined in the introduction which is the phase of the complex number ${\cal{X}}=\alpha \beta \gamma = \langle \psi_{{r}}\vert\psi_{{i}}\rangle \langle \psi_{{i}}\vert\psi_{{t}}\rangle \langle \psi_{{t}}\vert\psi_{{r}}\rangle$\cite{Berry1987}.  
Note that the two-terminal linear conductance across the junction (Eq.{\ref{T}}) depends upon two quantities : (a)   the  mismatch of density of states across the junction at the Fermi level, and (b)  ${\cal S}(\alpha,\beta,\gamma)$ which only depends on the spin texture mismatch of the dispersion across the junction at the Fermi level. The non-trivial aspect of the spin-texture mismatch lies in its dependence on $\Omega$ which is actually the only phase which influences the two terminal conductance. Next, we analyze the influence of $\Omega$ on the electrical transport.\\
{{\sl{\underline{Topological phase in $\phi=0$ case and the ${\mathbb Z}_2$ index} :}}} For $\phi=0$, the magnetic field points only along the $x$-direction while the SO field points along $z$-axis, hence the points representing incident, reflected and the transmitted spinors on the \bloch sphere are restricted to lie on a great circle contained in $x$-$z$ plane irrespective of the details of doping. As a consequence, the spherical triangle formed by connecting these three points (on the \bloch sphere) along the geodesic path\cite{sam} governed by details of  $\cal{X}$ has two distinct possibilities : either they encircle the centre of  \bloch sphere  once (call it $g_1$)  or it goes back and forth on a finite patch of the great circle (call it $g_2$) (see also \cite{pmcpc,lh}) without encircling the centre. The solid angle subtended by the closed curve $g_1$ at the centre is $2 \pi$ and that enclosed by $g_2$ is zero respectively. Using Pancharatnam's idea, we can immediately predict that $\Omega$, which is the phase of $\cal{X}$, should be equal to half the solid angle\cite{Berry1987} subtended by the geodesic triangles at the centre of \bloch sphere. This  implies that $g_1$ should correspond to $\Omega=\pi$ and $g_2$ should correspond  to $\Omega=0$ respectively, \ie ${\cal{X}}(g_1)$ is real and negative and ${\cal{X}}(g_2)$ is real and positive. Now, $g_1$ type of closed path which encircles the centre of \bloch sphere once are topologically distinct from that of $g_2$ type of path which do not encircle it at all and this fact manifests itself as sign of $\cal{X}$ in the transmission amplitude. It is worth noting that all closed paths restricted to the $x$-$z$ plane that are winding the origin of \bloch sphere even number of times are equivalent as the geometric phase associated with these paths are zero modulo $2 \pi$. Hence they are all equivalent to $g_2$. While all closed paths that are winding the origin of \bloch sphere odd number of times are equivalent as the geometric phase associated with these paths are $\pi$ modulo $2 \pi$ and hence they are equivalent to $g_1$. This facts imply that the conductance corresponding to the case of $g_1$ and $g_2$ can be distinguished in terms of a ${\mathbb Z}_2$ index $\nu$ given by 
\begin{equation}
(-1)^\nu=e^{ i \pi \nu} = \sgn\{ {\cal {X}}\} =\sgn \{\langle \psi_{{r}}\vert \psi_{{{i}}}\rangle \langle \psi_{{i}}\vert\psi_{{t}}\rangle \langle \psi_{{t}}\vert\psi_{{r}}\rangle \},
\end{equation} 
where $\nu$ takes two distinct inequivalent  values given by $\nu=0$ ( for $g_2$)  and $\nu=1$ ( for $g_1$). To develop an understanding of the physical situations that should correspond to ${\cal{X}}(g_1)$ and ${\cal{X}}(g_2)$, we define a parity of winding of the spin-texture for left and right moving electrons in the conduction and the valence band given by  
 \bea
 \lambda_{\eta,\epsilon}^{} =  \sgn \left[  \int_{0}^{(\epsilon \, \eta) \infty} \frac{\partial }{\partial k}\left( \theta_k^{\epsilon,\eta} \right) dk  \right]~,
 \label{}
\eea
where $\epsilon$ is defined in Eq.\ref{wf} and $\eta=\pm$ for right and left movers respectively. $\theta_k^{\epsilon,\eta}=\tan^{-1}(S_{k,z}^{\epsilon,\eta}/S_{k,x}^{\epsilon,\eta})$ where $S_{k,{z/x}}^{\epsilon,\eta}$ is the expectation value of the $z/x$ components of the spin operator. This is an interesting quantity as it tells us which way is the spin twisting as we move from $k=\pm \infty$ to $k=0$ in the conduction and the valence band. Note that as $k \to \infty$, the SO field dominates over external B-field and dictates the orientation of spin associated with the momentum mode while at $k=0$ the SO field vanishes and spin direction is dictated by the externally applied magnetic field. Essentially these  two facts decide the parity, $\lambda_{\eta,\epsilon}$. It is straight-forward to check that all junctions which have an opposite sign  of $\lambda$ for electrons of a given chirality (right or left movers) at the Fermi level on two sides of the junction always correspond to the ${\cal{X}}(g_1)$ (see Fig.\ref{fig2} in supplementary material) case.  Hence this situation corresponds to the $\nu=1$ case. This is the case for $p$-$n$ and $n$-$p$ junctions. On the other hand, if parity ($\lambda$) is same on two sides of the junction (\ie $n$-$n$ and $p$-$p$ junctions) then it  always correspond to ${\cal{X}}(g_2)$ and hence corresponds to $\nu=0$ case. This observation completes the topological classification of such junctions with negative value for the ${\mathbb Z}_2$ index for $p$-$n$, $n$-$p$ junctions and positive value for the ${\mathbb Z}_2$ index for $n$-$n$, $p$-$p$ junctions. This is one of the central results of this article.  Next we show that this topological phase ($\Omega$) becomes geometric as we turn on a finite $\phi$.\\
{\sl{\underline{Non-topological phase for $\phi \neq 0$ and junction dualities:}}} Now we consider a situation where the two sides of the junction are exposed to magnetic field pointing along two different directions in the $x$-$y$ plane  (as given in Eq.\ref{Bfield} ) which implies that the spin states associated with all momentum eigenstates belonging to the patch $-\infty \leq x < 0$ can be represented by points on \bloch sphere contained on a great circle lying in the  $x$-$z$ plane.  While all momentum eigenstates belonging to patch $0 < x \leq \infty $ will be represented on a great circle contained  in the \bloch sphere lying in the $n$-$z$ plane where $n=(B\cos\phi)\hat{x}+(B\sin\phi)\hat{y}$. From this fact it is clear that the incident, reflected and the transmitted spinor which goes into the construction of $\cal{X}$ can no longer be represented by three distinct points on the \bloch sphere which could be spotted on a single great circle. This in turn implies that half the solid angle subtended by the geodesic triangles formed by cyclic projection of these three states can never be equal to $2\pi$ (\ie $\Omega \neq \pi$)  and its value in general will depend on details of the position of these three states on the \bloch sphere hence rendering it non-topological or geometric (see inset of Fig.\ref{fig1}(b) for variation of $\Omega$) (also see Ref.\cite{pmcpv}). Actually the angle $\phi$ provides an efficient handle on this geometric phase which could be tuned to be topological in two limits, \ie for $\phi=0$ (studied above) and $\phi=\pi$ when all the states (incident, reflected and the transmitted spinor) once again lie on a single great circle on the \bloch sphere. The $\phi=\pi$ is an interesting case as it flips the signs of $\lambda$ hence resulting in switching between $g_1$ type of paths and $g_2$ type of paths. This amounts to saying that the $n$-$p$ junction for $\phi=0$ case is topologically equivalent to $n$-$n$ junction for the $\phi=\pi$ case and similarly $p$-$n$ junction for $\phi=0$  case is topologically equivalent to the $p$-$p$ junction for the $\phi=\pi$. Hence $\phi=0 \rightarrow \phi=\pi$ defines a duality (mapping between two topologically distinct junctions) between the $p$-$n$ and $n$-$n$ junctions and similarly between $n$-$p$ and $p$-$p$ junctions. Actually $\phi=0 \rightarrow \phi=\pi$ results in switching of two-terminal conductance between \{$p$-$n$,$n$-$n$\} junctions with \{$p$-$p$, $n$-$p$\} junctions respectively which is depicted in the plot in Fig.~\ref{fig1}(b). This is essentially a consequence of the fact that the spin textures of the valence and conduction band gets swapped as $\phi=0 \rightarrow \pi$, hence effectively converting a $n (p)$-doped region into a $p(n)$-doped region. \\ 
{{\sl{\underline{Proposed experimental protocol :}}}}  Above we have established a topological difference between the $n$-$n$ and $n$-$p$ junctions. To quantify this difference in terms of experimentally measurable quantity like conductance, we consider a situation where left gate ($V_{{\cal{G}}_1}=\delta_1$) is tuned to a n-type doping 
(\ie $\delta_1<0$ ) while the right gate ${V_{\cal {G}}}_2$ is kept flexible so that it can be tuned to a n-type (p-type) making the system a $n$-$n$ ($n$-$p$) junction by tuning the sign of ${V_{\cal{G}}}_2=\mp \delta_2$ ($\delta_2 > 0$) respectively. We also assume that the applied voltage bias is such that it is driving an electronic current from left to right.  Owing to the symmetry of the edge spectrum about the Dirac point, the value of $\rho_{i}^{} / \rho^{}_{t}$ (see Eq.\ref{T}) will be identical for a  $n$-$n$  and $n$-$p$ junction provided we keep  $V_{{\cal{G}}_1}=\delta_1$ fixed and  switch between the two types of junction only by changing the sign of ${V_{\cal{G}}}_2$ while keeping its magnitude fixed. 
Hence the difference in conductance (denoted by $\Delta G^\phi_{\delta_1,\delta_2}$) of a $n$-$n$ and $n$-$p$ junction for a given value of $V_{{\cal G}_1}$ common to both while $V_{{\cal G}_2}$ being different for the two junction only in sign leads to a quantity which depends linearly on the difference of ${\cal S} (\alpha,\beta,\gamma)$ (see Eq.\ref{T} ; call it $\Delta {\cal S}$). This is so because the dependence on $\rho^{}_{i}/ \rho^{}_{t}$ can be pulled out  of the expression as overall scale factor. Now, note that  $S(\alpha,\beta,\gamma)$ represents that part of conductance which depends only on the spin-texture mismatch at the junction and is solely responsible for the topological classification of that junction. Hence measurement of 
\bea
\Delta {\cal G}^{\phi}_{\delta_1,\delta_2}=({\cal G}^{\phi}_{\delta_1,\delta_2}-{\cal G}^{\phi}_{\delta_1,-\delta_2})\propto \Delta {\cal S}
 \label{delG}
\eea
could be viewed as a quantification of the topological difference between the $n$-$n$ and $n$-$p$ junction. Further more, due to the duality defined earlier between $n$-$n$ and $n$-$p$ junction, one expects the sign of $\Delta {\cal G}^{\phi}_{\delta_1,\delta_2}$ to flip keeping its magnitude fixed under  under the transformation $\phi=0 \rightarrow \phi=\pi$. This fact is depicted in Fig.\ref{fig1}(b) where $\Delta {\cal G}^{\phi}_{\delta_1,\delta_2}$ is plotted as a function of $\phi$. Hence an experimental measurement of $\Delta {\cal G}^{\phi}_{\delta_1,\delta_2}$ for $\phi=0$ and $\phi=\pi$ could provide a direct confirmation of the topological classification of these junctions presented in this article. Also the variation of conductance as function of strength of magnetic field-$\vec{B}$ with uniform direction and a symmetric gate voltage configuration (given $\delta_1$ and opposite sign for given $\delta_2$ for $n$-$n$ and $n$-$p$ junction) shows asymmetry in $n$-$n$ and $n$-$p$ conductance which is induced by spin texture mismatch at the junction (see inset of Fig.~\ref{fig1}(a). This asymmetry can also be seen for symmetric gate voltage configuration when conductance is plotted as a function of gate voltage $\delta_2$  while magnetic field is kept uniform and its strength is held fixed (see Fig.~\ref{fig1}(a)).\\
{{\sl{\underline{Interaction correlation to conductance:}}}} Inclusion of electron-electron interaction on the edge leads to universal  Luttinger liquid phase known for interacting 1-D electron liquid\cite{Kane1992}.  As the spin Hall edge state is helical in nature the Luttinger liquid phase of such edge states lead to the helical Luttinger liquid phase \cite{Wu2006,Teo2009,llan2012}. Two important parameters which parametrizes the helical Luttinger liquid theory are the effective interaction parameter $K$ and the renormalized Fermi-velocity $v$ which are connected by Galilean invariance as $K  v = 1 $\cite{Maslov2005}.  To proceed further we first linearize the spectrum around the Fermi level which could either be in the conduction or the valence band. Then the corresponding creation operator for electrons living inside the linearized energy window could be written as
 \bea
 \psi_{\epsilon}(x) = \psi_{{R, \epsilon}} e_{}^{i  \epsilon  k_F x}  a^{}
 _{R,\epsilon}(x) + \psi_{{L, \epsilon}} e^{-i  \epsilon k_F x}  a^{}_{L,\epsilon}(x),
 \label{K}
\eea
where $\psi_{{R/L,\epsilon}}$ are spinors evaluated at the Fermi energy for right (R) and left (L) movers and they can be imported directly from Eq.\ref{wf}, $k_F$ represents the Fermi momentum and $a^{}_{R/L,\epsilon}(x)$ represents slow chiral degrees of freedom which could be bosonized using standard bosonization technique\cite{Delft1998}. Following Ref.\onlinecite{Suhas2014} for deriving an expression for Luttinger parameter for doped helical Luttinger liquid and following Ref.\onlinecite{Maslov2005} for carefully implementing the requirement of Galilean invariance,  it is straight-forward to obtain the Luttinger parameter for the conduction($\epsilon=1$) and valence($\epsilon=-1$) band as  
\bea
 K_{\epsilon} =  \left\{ { \left(1+\frac{V(0)}{\pi v_F^{}}\right) -\frac{V(2 k_F^{}) |\langle \psi_{{L,\epsilon}}  \vert \psi_{{R,\epsilon}}\rangle|^2}{\pi v_F}}\right\}^{-1/2}~.
 \label{K}
\eea
where $V(0)$ and $V(2 k_F)$ are the Fourier transforms of screened Coulomb potential at zero and $2 k_F$ momenta. From Eq.\ref{wf} it can be checked that the quantity  $|\langle \psi^{}_{{L,\epsilon}}  \vert \psi_{{R,\epsilon}}^{} \rangle|$ is identical for the conduction and the valence band provided we compare situations with identical doping with respect to the zero doping situation (\ie with respect to situation corresponding to the Fermi level lying in the middle of the gap).  This fact provides a clear indication that the both for $n$-$n$ and $n$-$p$ junction, the effect due to electron-electron interaction are expected to be same as long the the symmetry of the spectrum about the  middle of the gap stays intact.\\
{{\sl{\underline{Experimental feasibility study :} }}}  \\
{{\sl{{(i) Energy and length scales-} }}}   A typical 2D system which hosts a helical 1D state as an edge state is the quantum spin Hall (QSH) state\cite{KaneReview2010,zhang2011} which was experimentally observed in HgTe/CdTe Quantum Wells\cite{Konig2008}. As our proposed experiment requires application of magnetic field on the 1-D helical state which leads to breaking of TRS in general, an estimate of hierarchy of energy scales, which will protect the topological gap in the bulk but will allow for minimal breaking of TRS on the edge leading to a gapped edge spectrum is desirable. Experiments by  K$\ddot{\rm o}$nig et al.\cite{Konig2008}  reported that the QSH state is destroyed by a magnetic field of magnitude $0.7 T$ (in-plane) and $28 mT$ (out of plane). As the SO field for the edge state point perpendicular to the plane of the 2-D quantum wells, our proposal requires application of an in plane field. Hence, an in-plane field of the order of $50\, \text{mT}$ can be considered safe as it will hardly distort the bulk QSH state but will surely open up a gap in the edge spectrum. A $50\, \text{mT}$  field amounts  to  gap of $145 \mu eV$  in the edge spectrum (if we assume a Land\'e-g factor of $50$ which is the g-factor for the bulk HgTe) and it further amounts to an equivalent temperature of  $1.7 K$ while the experiments were done at temperatures of the order of $30 \text{mK}$\cite{Konig2008}. Hence this gap can be readily resolved in present day experimental setups which makes our proposal quite feasible. 
Experiment of Karmakar et al.\cite{Biswajit2011} suggests our proposal could be implemented on the QSH edge by using a nano-magnetic array. 
Also, experimentally observed elastic mean free path in these edge states are $\sim$ $1\,\mu m$\cite{Daumer2003,Konig2008}. As our calculations were performed assuming ballistic limit, a device of length of a few micrometers could be optimal for implementing our proposal. {{\sl{{(ii) Step function form of gate voltage and B field -} }}} As our theoretical analysis assumes a step function like variation of gate voltage $V_{{{\cal G}}}$  and magnetic field  $\vec{B}$ across the $p$-$n$ junction, it is important to understand, variation over what length scales \cite{PhysRevB.87.100506,2015EPJB...88...62K} can be approximated as step function. From Eq.(\ref{H}) we can idetify the length scales associated with the change in chemical potential and the magnetic field across the junction as $l_{gate}=({v_{F}\hbar}) /({e\vert \pm V_{{{\cal G}}_2}-V_{{{\cal G}}_1}\vert})$ and  $l_{B}=({v_{F}\hbar})/(g\mu_{B}\vert \vec{B_{1}}-\vec{B_{2}}\vert)$. When $\vec{B_{1}}$ and $\vec{B_{2}}$ are diagrammatically opposite but their magnitudes are same (say, $B$) then the factor $\vert \vec{B_{1}}-\vec{B_{2}}\vert= 2B$. A linear variation of gate voltage or the $\vec{B}$ which happens over length $l << l_{gate},l_{B}$ can be taken to be of step function form as far as conductance is concerned (refer supplementary material)\cite{PhysRevLett.115.136804}.  {{\sl{{(iii) Symmetry of edge spectrum -} }}} Note that symmetry of the edge spectrum about its Dirac point is an important ingredient in our proposal. In general the Dirac point of edge spectrum may not lie in the middle of the bulk gap and  additionally, the chemical potentials of the material may not lie at the Dirac point. Hence to implement our proposal for the $p$-$n$ junction, the edge chemical potential should be initialized to the Dirac point by application of a global electrostatic gate. Further electrostatic gate voltages can be used with respect to this global gate voltage to tune the position of the edge chemical potential separately on the two sides of the junction.\\
{{\sl{\underline{Conclusion and outlook:} }}} 
In an insightful paper by Kane and Mele\cite{Kane2005} it was shown that the quantum spin Hall phase can be classified in terms of a time reversal symmetry protected ${\mathbb Z}_2$ index which hosts helical edge state carrying a net equilibrium spin current. In this article we have shown that non-equilibrium charge transport (transport driven by bias) across a $p$-$n$ junction realized on these helical edge states can also have ${\mathbb Z}_2$ topological classification. In our case the ${\mathbb Z}_2$ topology, which is directly related to helical nature of the edge spectrum, stays protected as long as the $p$-$n$ junction is exposed to an uniaxial magnetic field. It is worth noting, in contrast to the ongoing efforts to use topology for classifying phases of condensed matter systems, in this article we present a topological classification of measurable quantities like conductance for a device element such as $p$-$n$ junction. This view point could lead to innovative protocols for device manipulation as demonstrated in the work.\\
{{\sl{\underline{ Acknowledgments:}}} It is a pleasure to thank Jens Bardarson, Suhas Gangadharaiah for discussion,  Laurens W. Molenkamp and Andrea Young for discussion and critical comments on experimental feasibility of the proposed measurement. We thank Yin-Chen-He for critical reading of the manuscript. DW would like to thank Krishnu Roy Chowdhury $\&$ Santhust for helping with Python Programming. DW acknowledges University Teaching Assistantship programme offered by the University of Delhi and PM acknowledges support from the University Grants Commission under the second phase of University with Potential of Excellence at Jawaharlal Nehru University. SD acknowledges support from University of Delhi in the form of a research grant (RC/2014/6820).
\vskip -1.5em
\bibliographystyle{apsrev}
\bibliography{references}

\begin{thebibliography}{30}
\expandafter\ifx\csname natexlab\endcsname\relax\def\natexlab#1{#1}\fi
\expandafter\ifx\csname bibnamefont\endcsname\relax
  \def\bibnamefont#1{#1}\fi
\expandafter\ifx\csname bibfnamefont\endcsname\relax
  \def\bibfnamefont#1{#1}\fi
\expandafter\ifx\csname citenamefont\endcsname\relax
  \def\citenamefont#1{#1}\fi
\expandafter\ifx\csname url\endcsname\relax
  \def\url#1{\texttt{#1}}\fi
\expandafter\ifx\csname urlprefix\endcsname\relax\def\urlprefix{URL }\fi
\providecommand{\bibinfo}[2]{#2}
\providecommand{\eprint}[2][]{\url{#2}}

\bibitem[{\citenamefont{Berry}(1984)}]{Berry:1984jv}
\bibinfo{author}{\bibfnamefont{M.~V.} \bibnamefont{Berry}},
  \bibinfo{journal}{Proc. Roy. Soc. Lond.} \textbf{\bibinfo{volume}{A392}},
  \bibinfo{pages}{45} (\bibinfo{year}{1984}).

\bibitem[{\citenamefont{Shapere and Wilczek}(1989)}]{shaperebook}
\bibinfo{author}{\bibfnamefont{A.}~\bibnamefont{Shapere}} \bibnamefont{and}
  \bibinfo{author}{\bibfnamefont{F.}~\bibnamefont{Wilczek}},
  \emph{\bibinfo{title}{Geometric Phases in Physics}}
  (\bibinfo{publisher}{World Scientific, Singapore}, \bibinfo{year}{1989}).

\bibitem[{\citenamefont{Nakagawa}(1987)}]{nakagawa}
\bibinfo{author}{\bibfnamefont{N.}~\bibnamefont{Nakagawa}},
  \bibinfo{journal}{Ann. Phys.} \textbf{\bibinfo{volume}{179}},
  \bibinfo{pages}{145} (\bibinfo{year}{1987}).

\bibitem[{\citenamefont{Aharonov and Anandan}(1987)}]{Aharonov:1987gg}
\bibinfo{author}{\bibfnamefont{Y.}~\bibnamefont{Aharonov}} \bibnamefont{and}
  \bibinfo{author}{\bibfnamefont{J.}~\bibnamefont{Anandan}},
  \bibinfo{journal}{Phys. Rev. Lett.} \textbf{\bibinfo{volume}{58}},
  \bibinfo{pages}{1593} (\bibinfo{year}{1987}).

\bibitem[{\citenamefont{Samuel and Bhandari}(1988)}]{sam}
\bibinfo{author}{\bibfnamefont{J.}~\bibnamefont{Samuel}} \bibnamefont{and}
  \bibinfo{author}{\bibfnamefont{R.}~\bibnamefont{Bhandari}},
  \bibinfo{journal}{Phys. Rev. Lett.} \textbf{\bibinfo{volume}{60}},
  \bibinfo{pages}{2339} (\bibinfo{year}{1988}).

\bibitem[{\citenamefont{Ramaseshan and Nityananda}(1986)}]{ramaseshan1986}
\bibinfo{author}{\bibfnamefont{S.}~\bibnamefont{Ramaseshan}} \bibnamefont{and}
  \bibinfo{author}{\bibfnamefont{R.}~\bibnamefont{Nityananda}},
  \bibinfo{journal}{Curr. Sci.} \textbf{\bibinfo{volume}{55}},
  \bibinfo{pages}{1225} (\bibinfo{year}{1986}).

\bibitem[{\citenamefont{Berry}(1987)}]{Berry1987}
\bibinfo{author}{\bibfnamefont{M.}~\bibnamefont{Berry}},
  \bibinfo{journal}{Journal of Modern Optics} \textbf{\bibinfo{volume}{34}},
  \bibinfo{pages}{1401} (\bibinfo{year}{1987}), ISSN \bibinfo{issn}{0950-0340}.

\bibitem[{\citenamefont{Pancharatnam}(1956)}]{Pancharatnam:1956}
\bibinfo{author}{\bibfnamefont{S.}~\bibnamefont{Pancharatnam}},
  \bibinfo{journal}{Proc. Ind. Acad. Sci.} \textbf{\bibinfo{volume}{A44}},
  \bibinfo{pages}{247} (\bibinfo{year}{1956}).

\bibitem[{\citenamefont{Wu et~al.}(2006)\citenamefont{Wu, Bernevig, and
  Zhang}}]{Wu2006}
\bibinfo{author}{\bibfnamefont{C.}~\bibnamefont{Wu}},
  \bibinfo{author}{\bibfnamefont{B.}~\bibnamefont{Bernevig}}, \bibnamefont{and}
  \bibinfo{author}{\bibfnamefont{S.-C.} \bibnamefont{Zhang}},
  \bibinfo{journal}{Phys. Rev. Lett.} \textbf{\bibinfo{volume}{96}},
  \bibinfo{pages}{106401} (\bibinfo{year}{2006}).

\bibitem[{\citenamefont{{Yokoyama} and {Murakami}}(2014)}]{yoko2014}
\bibinfo{author}{\bibfnamefont{T.}~\bibnamefont{{Yokoyama}}} \bibnamefont{and}
  \bibinfo{author}{\bibfnamefont{S.}~\bibnamefont{{Murakami}}},
  \bibinfo{journal}{Physica E Low-Dimensional Systems and Nanostructures}
  \textbf{\bibinfo{volume}{55}}, \bibinfo{pages}{1} (\bibinfo{year}{2014}).

\bibitem[{\citenamefont{K\"{o}nig et~al.}(2007)\citenamefont{K\"{o}nig,
  Wiedmann, Br\"{u}ne, Roth, Buhmann, Molenkamp, Qi, and Zhang}}]{Konig2007}
\bibinfo{author}{\bibfnamefont{M.}~\bibnamefont{K\"{o}nig}},
  \bibinfo{author}{\bibfnamefont{S.}~\bibnamefont{Wiedmann}},
  \bibinfo{author}{\bibfnamefont{C.}~\bibnamefont{Br\"{u}ne}},
  \bibinfo{author}{\bibfnamefont{A.}~\bibnamefont{Roth}},
  \bibinfo{author}{\bibfnamefont{H.}~\bibnamefont{Buhmann}},
  \bibinfo{author}{\bibfnamefont{L.~W.} \bibnamefont{Molenkamp}},
  \bibinfo{author}{\bibfnamefont{X.-L.} \bibnamefont{Qi}}, \bibnamefont{and}
  \bibinfo{author}{\bibfnamefont{S.-C.} \bibnamefont{Zhang}},
  \bibinfo{journal}{Science (New York, N.Y.)} \textbf{\bibinfo{volume}{318}},
  \bibinfo{pages}{766} (\bibinfo{year}{2007}), ISSN \bibinfo{issn}{1095-9203}.

\bibitem[{\citenamefont{{Soori} et~al.}(2012)\citenamefont{{Soori}, {Das}, and
  {Rao}}}]{Soori2012}
\bibinfo{author}{\bibfnamefont{A.}~\bibnamefont{{Soori}}},
  \bibinfo{author}{\bibfnamefont{S.}~\bibnamefont{{Das}}}, \bibnamefont{and}
  \bibinfo{author}{\bibfnamefont{S.}~\bibnamefont{{Rao}}},
  \bibinfo{journal}{\prb} \textbf{\bibinfo{volume}{86}}, \bibinfo{eid}{125312}
  (\bibinfo{year}{2012}).

\bibitem[{\citenamefont{Mehta}(2009{\natexlab{a}})}]{pmcpc}
\bibinfo{author}{\bibfnamefont{P.}~\bibnamefont{Mehta}},
  \bibinfo{journal}{Phys. Rev. D} \textbf{\bibinfo{volume}{79}},
  \bibinfo{eid}{096013} (\bibinfo{year}{2009}{\natexlab{a}}).

\bibitem[{\citenamefont{Longuet-Higgins
  et~al.}(1958)\citenamefont{Longuet-Higgins, Opik, Pryce, and Sack}}]{lh}
\bibinfo{author}{\bibfnamefont{H.~C.} \bibnamefont{Longuet-Higgins}},
  \bibinfo{author}{\bibfnamefont{U.}~\bibnamefont{Opik}},
  \bibinfo{author}{\bibfnamefont{M.~H.~L.} \bibnamefont{Pryce}},
  \bibnamefont{and} \bibinfo{author}{\bibfnamefont{R.~A.} \bibnamefont{Sack}},
  \bibinfo{journal}{Proc. Roy. Soc. Lond.} \textbf{\bibinfo{volume}{A244}},
  \bibinfo{pages}{1} (\bibinfo{year}{1958}).

\bibitem[{\citenamefont{Mehta}(2009{\natexlab{b}})}]{pmcpv}
\bibinfo{author}{\bibfnamefont{P.}~\bibnamefont{Mehta}}
  (\bibinfo{year}{2009}{\natexlab{b}}), \eprint{0907.0562}.

\bibitem[{\citenamefont{Kane and Fisher}(1992)}]{Kane1992}
\bibinfo{author}{\bibfnamefont{C.~L.} \bibnamefont{Kane}} \bibnamefont{and}
  \bibinfo{author}{\bibfnamefont{M.~P.~A.} \bibnamefont{Fisher}},
  \bibinfo{journal}{Phys. Rev. B} \textbf{\bibinfo{volume}{46}},
  \bibinfo{pages}{15233} (\bibinfo{year}{1992}).

\bibitem[{\citenamefont{Teo and Kane}(2009)}]{Teo2009}
\bibinfo{author}{\bibfnamefont{J.~C.~Y.} \bibnamefont{Teo}} \bibnamefont{and}
  \bibinfo{author}{\bibfnamefont{C.~L.} \bibnamefont{Kane}},
  \bibinfo{journal}{Phys. Rev. B} \textbf{\bibinfo{volume}{79}},
  \bibinfo{pages}{235321} (\bibinfo{year}{2009}).

\bibitem[{\citenamefont{{Ilan} et~al.}(2012)\citenamefont{{Ilan}, {Cayssol},
  {Bardarson}, and {Moore}}}]{llan2012}
\bibinfo{author}{\bibfnamefont{R.}~\bibnamefont{{Ilan}}},
  \bibinfo{author}{\bibfnamefont{J.}~\bibnamefont{{Cayssol}}},
  \bibinfo{author}{\bibfnamefont{J.~H.} \bibnamefont{{Bardarson}}},
  \bibnamefont{and} \bibinfo{author}{\bibfnamefont{J.~E.}
  \bibnamefont{{Moore}}}, \bibinfo{journal}{Physical Review Letters}
  \textbf{\bibinfo{volume}{109}}, \bibinfo{eid}{216602} (\bibinfo{year}{2012}).

\bibitem[{\citenamefont{{Maslov}}(2005)}]{Maslov2005}
\bibinfo{author}{\bibfnamefont{D.~L.} \bibnamefont{{Maslov}}},
  \bibinfo{journal}{eprint arXiv:cond-mat/0506035}  (\bibinfo{year}{2005}),
  \eprint{cond-mat/0506035}.

\bibitem[{\citenamefont{{von Delft} and {Schoeller}}(1998)}]{Delft1998}
\bibinfo{author}{\bibfnamefont{J.}~\bibnamefont{{von Delft}}} \bibnamefont{and}
  \bibinfo{author}{\bibfnamefont{H.}~\bibnamefont{{Schoeller}}},
  \bibinfo{journal}{Annalen der Physik} \textbf{\bibinfo{volume}{7}},
  \bibinfo{pages}{225} (\bibinfo{year}{1998}).

\bibitem[{\citenamefont{{Gangadharaiah}
  et~al.}(2014)\citenamefont{{Gangadharaiah}, {Schmidt}, and
  {Loss}}}]{Suhas2014}
\bibinfo{author}{\bibfnamefont{S.}~\bibnamefont{{Gangadharaiah}}},
  \bibinfo{author}{\bibfnamefont{T.~L.} \bibnamefont{{Schmidt}}},
  \bibnamefont{and} \bibinfo{author}{\bibfnamefont{D.}~\bibnamefont{{Loss}}},
  \bibinfo{journal}{\prb} \textbf{\bibinfo{volume}{89}}, \bibinfo{eid}{035131}
  (\bibinfo{year}{2014}).

\bibitem[{\citenamefont{Hasan and Kane}(2010)}]{KaneReview2010}
\bibinfo{author}{\bibfnamefont{M.~Z.} \bibnamefont{Hasan}} \bibnamefont{and}
  \bibinfo{author}{\bibfnamefont{C.~L.} \bibnamefont{Kane}},
  \bibinfo{journal}{Rev. Mod. Phys.} \textbf{\bibinfo{volume}{82}},
  \bibinfo{pages}{3045} (\bibinfo{year}{2010}).

\bibitem[{\citenamefont{Qi and Zhang}(2011)}]{zhang2011}
\bibinfo{author}{\bibfnamefont{X.-L.} \bibnamefont{Qi}} \bibnamefont{and}
  \bibinfo{author}{\bibfnamefont{S.-C.} \bibnamefont{Zhang}},
  \bibinfo{journal}{Rev. Mod. Phys.} \textbf{\bibinfo{volume}{83}},
  \bibinfo{pages}{1057} (\bibinfo{year}{2011}).

\bibitem[{\citenamefont{K\"{o}nig et~al.}(2008)\citenamefont{K\"{o}nig,
  Buhmann, {W. Molenkamp}, Hughes, Liu, Qi, and Zhang}}]{Konig2008}
\bibinfo{author}{\bibfnamefont{M.}~\bibnamefont{K\"{o}nig}},
  \bibinfo{author}{\bibfnamefont{H.}~\bibnamefont{Buhmann}},
  \bibinfo{author}{\bibfnamefont{L.}~\bibnamefont{{W. Molenkamp}}},
  \bibinfo{author}{\bibfnamefont{T.}~\bibnamefont{Hughes}},
  \bibinfo{author}{\bibfnamefont{C.-X.} \bibnamefont{Liu}},
  \bibinfo{author}{\bibfnamefont{X.-L.} \bibnamefont{Qi}}, \bibnamefont{and}
  \bibinfo{author}{\bibfnamefont{S.-C.} \bibnamefont{Zhang}},
  \bibinfo{journal}{JPSJ} \textbf{\bibinfo{volume}{77}},
  \bibinfo{pages}{031007} (\bibinfo{year}{2008}).

\bibitem[{\citenamefont{{Karmakar} et~al.}(2011)\citenamefont{{Karmakar},
  {Venturelli}, {Chirolli}, {Taddei}, {Giovannetti}, {Fazio}, {Roddaro},
  {Biasiol}, {Sorba}, {Pellegrini} et~al.}}]{Biswajit2011}
\bibinfo{author}{\bibfnamefont{B.}~\bibnamefont{{Karmakar}}},
  \bibinfo{author}{\bibfnamefont{D.}~\bibnamefont{{Venturelli}}},
  \bibinfo{author}{\bibfnamefont{L.}~\bibnamefont{{Chirolli}}},
  \bibinfo{author}{\bibfnamefont{F.}~\bibnamefont{{Taddei}}},
  \bibinfo{author}{\bibfnamefont{V.}~\bibnamefont{{Giovannetti}}},
  \bibinfo{author}{\bibfnamefont{R.}~\bibnamefont{{Fazio}}},
  \bibinfo{author}{\bibfnamefont{S.}~\bibnamefont{{Roddaro}}},
  \bibinfo{author}{\bibfnamefont{G.}~\bibnamefont{{Biasiol}}},
  \bibinfo{author}{\bibfnamefont{L.}~\bibnamefont{{Sorba}}},
  \bibinfo{author}{\bibfnamefont{V.}~\bibnamefont{{Pellegrini}}},
  \bibnamefont{et~al.}, \bibinfo{journal}{Physical Review Letters}
  \textbf{\bibinfo{volume}{107}}, \bibinfo{eid}{236804} (\bibinfo{year}{2011}),
  \eprint{1106.3965}.

\bibitem[{\citenamefont{Daumer et~al.}(2003)\citenamefont{Daumer, Golombek,
  Gbordzoe, Novik, Hock, Becker, Buhmann, and Molenkamp}}]{Daumer2003}
\bibinfo{author}{\bibfnamefont{V.}~\bibnamefont{Daumer}},
  \bibinfo{author}{\bibfnamefont{I.}~\bibnamefont{Golombek}},
  \bibinfo{author}{\bibfnamefont{M.}~\bibnamefont{Gbordzoe}},
  \bibinfo{author}{\bibfnamefont{E.~G.} \bibnamefont{Novik}},
  \bibinfo{author}{\bibfnamefont{V.}~\bibnamefont{Hock}},
  \bibinfo{author}{\bibfnamefont{C.~R.} \bibnamefont{Becker}},
  \bibinfo{author}{\bibfnamefont{H.}~\bibnamefont{Buhmann}}, \bibnamefont{and}
  \bibinfo{author}{\bibfnamefont{L.~W.} \bibnamefont{Molenkamp}},
  \bibinfo{journal}{Applied Physics Letters} p. \bibinfo{pages}{1376}
  (\bibinfo{year}{2003}), ISSN \bibinfo{issn}{00036951}.

\bibitem[{\citenamefont{Ojanen}(2013)}]{PhysRevB.87.100506}
\bibinfo{author}{\bibfnamefont{T.}~\bibnamefont{Ojanen}},
  \bibinfo{journal}{Phys. Rev. B} \textbf{\bibinfo{volume}{87}},
  \bibinfo{pages}{100506} (\bibinfo{year}{2013}).

\bibitem[{\citenamefont{{Klinovaja} and {Loss}}(2015)}]{2015EPJB...88...62K}
\bibinfo{author}{\bibfnamefont{J.}~\bibnamefont{{Klinovaja}}} \bibnamefont{and}
  \bibinfo{author}{\bibfnamefont{D.}~\bibnamefont{{Loss}}},
  \bibinfo{journal}{European Physical Journal B} \textbf{\bibinfo{volume}{88}},
  \bibinfo{eid}{62} (\bibinfo{year}{2015}), \eprint{1408.3366}.

\bibitem[{\citenamefont{Li et~al.}(2015)\citenamefont{Li, Wang, Fu, Du,
  Schreiber, Mu, Liu, Sullivan, Cs\'athy, Lin et~al.}}]{PhysRevLett.115.136804}
\bibinfo{author}{\bibfnamefont{T.}~\bibnamefont{Li}},
  \bibinfo{author}{\bibfnamefont{P.}~\bibnamefont{Wang}},
  \bibinfo{author}{\bibfnamefont{H.}~\bibnamefont{Fu}},
  \bibinfo{author}{\bibfnamefont{L.}~\bibnamefont{Du}},
  \bibinfo{author}{\bibfnamefont{K.~A.} \bibnamefont{Schreiber}},
  \bibinfo{author}{\bibfnamefont{X.}~\bibnamefont{Mu}},
  \bibinfo{author}{\bibfnamefont{X.}~\bibnamefont{Liu}},
  \bibinfo{author}{\bibfnamefont{G.}~\bibnamefont{Sullivan}},
  \bibinfo{author}{\bibfnamefont{G.~A.} \bibnamefont{Cs\'athy}},
  \bibinfo{author}{\bibfnamefont{X.}~\bibnamefont{Lin}}, \bibnamefont{et~al.},
  \bibinfo{journal}{Phys. Rev. Lett.} \textbf{\bibinfo{volume}{115}},
  \bibinfo{pages}{136804} (\bibinfo{year}{2015}).

\bibitem[{\citenamefont{{Kane} and {Mele}}(2005)}]{Kane2005}
\bibinfo{author}{\bibfnamefont{C.~L.} \bibnamefont{{Kane}}} \bibnamefont{and}
  \bibinfo{author}{\bibfnamefont{E.~J.} \bibnamefont{{Mele}}},
  \bibinfo{journal}{Physical Review Letters} \textbf{\bibinfo{volume}{95}},
  \bibinfo{eid}{146802} (\bibinfo{year}{2005}).

\end{thebibliography}
\pagebreak
\begin{center}
\textbf{Supplementary Material : Numerical Calculation of Conductance}
\end{center}
\begin{figure}[htb!]
\includegraphics[width=1.2\linewidth, height=.38\linewidth]{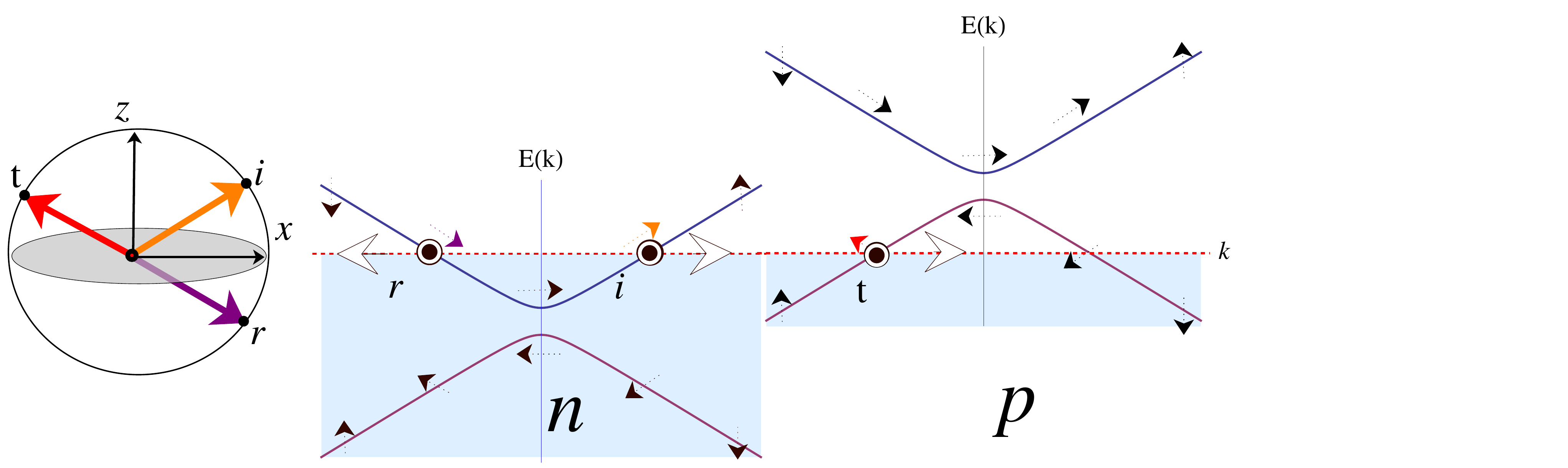}
\caption{The right panel shows the dispersion for $n$ doped ($x<0$)  and $p$ doped ($x>0$) sides of a $n$-$p$ junction where the black arrows show the spin texture of the dispersion. The left panel shows the spin states of incident ($i$, orange), reflected ($r$, magenta) and transmitted ($t$, red) electron wave functions at the Fermi level represented as three dots placed on the great circle contained in $x$-$z$ plane on the \bloch sphere. }
\label{fig2}
\end{figure}
To analyse the validity of our model considered in eqn.(\ref{H}) where we have assumed that both the gate voltage as well as the applied in-plane magnetic field change abruptly at the junction, we will now study a model with a junction extended over length $l$ where the applied electric and magnetic fields are connected to their assymptotic values on the two sides of the junction via a linear interpolation over a  length $l$. \\\\We can identify length scales in our model directly from the Hamiltonian (see Eq.(1)). The length scales associated with the change in chemical potential and the magnetic field across the junction are $l_{gate}=({v_{F}\hbar}) /({e\vert \pm V_{{{\cal G}}_2}-V_{{{\cal G}}_1}\vert})$ and  $l_{B}=({v_{F}\hbar})/(g\mu_{B}\vert \vec{B_{1}}-\vec{B_{2}}\vert)$. When $\vec{B_{1}}$ and $\vec{B_{2}}$ are diagrammatically opposite assuming their magnitudes are same (say, $B$) then the factor $\vert \vec{B_{1}}-\vec{B_{2}}\vert= 2B$ . If the scale over which variation at the junction happens is shorter than the length scales $l_{gate}$ and $l_{B}$ determined by the difference in assymptotic values of the chemical potential and the magnetic field respectively on the two sides of the junction then, our assumption of abrupt junction faithfully describes the situation as shown below through numerical analysis. \\\\
Now, our model with an extended junction can be divided into three regions namely, I ($-\infty<x<-l/2$), II ($-l/2\leq x\leq l/2$) and III ($l/2<x<\infty$). We further divide the middle region \emph{i.e.} region of extended junction into $n$ small patches such that
the Hamiltonian in the three regions  can be written as -
\begin{eqnarray}
{\cal H}_{I} = - i \hbar v_{F} \sigma_{z}\partial_{x} + g\mu_{B}\left(\vec{S} \cdot \vec{B_{1}}(x,y)\right) 
   + eV_{{{\cal G}}_1}(x) \nonumber\\
{\cal H}_{II,n} = - i \hbar v_{F} \sigma_{z}\partial_{x} + g\mu_{B}\left(\vec{S} \cdot \vec{B_{n}}(x,y)  \right) 
   + e V_{{{\cal G}}_{n}}(x) \nonumber \\
{\cal H}_{III} = - i \hbar v_{F} \sigma_{z}\partial_{x} + g\mu_{B}\left(\vec{S} \cdot \vec{B_{2}}(x,y)\right) 
   + eV_{{{\cal G}}_2}(x)  \nonumber
\end{eqnarray}
Where, $\vec{B_{1}}=B(\cos{\phi_{1}}\hat{x}+\sin{\phi_{1}}\hat{y})$ ; $\vec{B_{2}}=B(\cos{\phi_{2}}\hat{x}+\sin{\phi_{2}}\hat{y})$ ;
$\vec{B_{n}} = B(\cos{\phi_{n}}\hat{x}+\sin{\phi_{n}}\hat{y})$ ; $\phi_{n}=\phi_{1}+(n-1)\Delta \phi$ ; $\Delta \phi = (\phi_{2}-\phi_{1})/n$ and $ V_{{{\cal G}}_{n}}=V_{{{\cal G}}_1}-(n-1)\Delta V_{{{\cal G}}}$ with $\Delta V_{{{\cal G}}}=(\vert V_{{{\cal G}}_2}-V_{{{\cal G}}_1}\vert)/n$.
 \\The wavefunction matching at $x=-l/2$ can be expressed as -
\begin{equation}
\psi_{i } e^{\iota k_{i} d_{1} } + r \psi_{r } e^{-\iota k_{i } d_{1}}=A_{1 } \Psi_{1}^{+} e^{\iota k_{1} d_{1}}+B_{1} \Psi_{1}^{-} e^{-\iota k_{1} d_{1}}  \nonumber
\end{equation}
where, $r$ is the reflection amplitude, $d_{1}=-l/2$, $\psi_{i}$ and $\psi_{r}$ are the incident and the reflected spinors respectively (see Eq.\ref{wf}).
$A_{1}$ and $B_{1}$ are the amplitudes associated with spinors $\Psi_{1}^{+}$ and $\Psi_{1}^{-}$ corresponding to Hamiltonian ${\cal H}_{II,n=1}$. The spinors in the region $l$ for the case of $n$-$n$ junction can be expressed as -
\begin{equation}
\Psi_{n}^{\pm}= (1 / N_{\pm}) 
\left[ \begin{array}{cc} ( \pm k_{n}^{} +  \sqrt{k_{n}^{2} + b^{2}_{}})/(b e^{i \phi_{n}})   & 1  \end{array}  \right]^{T} \nonumber
\end{equation}
while in the case of $n$-$p$ junction, the spinors in the region $d_{n}>0$ are replaced appropriately with -
\begin{equation}
\Psi_{n}^{\pm}= (1 / N_{\pm}) 
\left[ \begin{array}{cc} ( \mp k_{n}^{} -  \sqrt{k_{n}^{2} + b^{2}_{}})/(b e^{i \phi_{n}})   & 1  \end{array}  \right]^{T} \nonumber
\end{equation}
with, $k_{n}=\sqrt{(V_{{{\cal G}}_{n}}/v_{F})^{2}+b^{2}}$ and $N_{\pm}$ is the normalization constant.
\\ In the extended junction region $-l/2 < x < l/2$ the wavefunction at each $(n-1)^{th}$ step is matched to wavefunction at $n^{th}$ step as -
\begin{equation}
\begin{split}
A_{(n-1) } \Psi_{(n-1) }^{+} e^{\iota k_{(n-1)} d_{n}}+B_{(n-1)} \Psi_{(n-1)}^{-} e^{-\iota k_{(n-1)} d_{n}}\\=A_{n } \Psi_{n^{ } }^{+} e^{\iota k_{n^{ }} d_{n}}+B_{n^{ }} \Psi_{n^{ }}^{-} e^{-\iota k_{n^{ }} d_{n}}  \nonumber
\end{split}
\end{equation}
with $d_{n} = -l/2+(n-1)\Delta d$ ; $\Delta d=l/n$.
$A_{n}$ and $B_{n}$ are the amplitudes associated to the spinors $\Psi_{n}^{\pm}$ in region II in the $n^{th}$ patch. 
\\And, at $d_{n+1}= l/2$, wavefunction matching is -
\begin{equation}
A_{n } \Psi_{n }^{+} e^{\iota k_{n} d_{n+1}}+B_{n} \Psi_{n}^{-} e^{-\iota k_{n} d_{n+1}}= t \psi_{t} e^{\iota k_{t} d_{n+1}} \nonumber
\end{equation}
where $t$ is the transmission amplitude of the transmitted spinor $\psi_{t}$ as defined in eqn.(\ref{wf}).
\begin{figure}[htb!]
\includegraphics[width=1.0\linewidth]{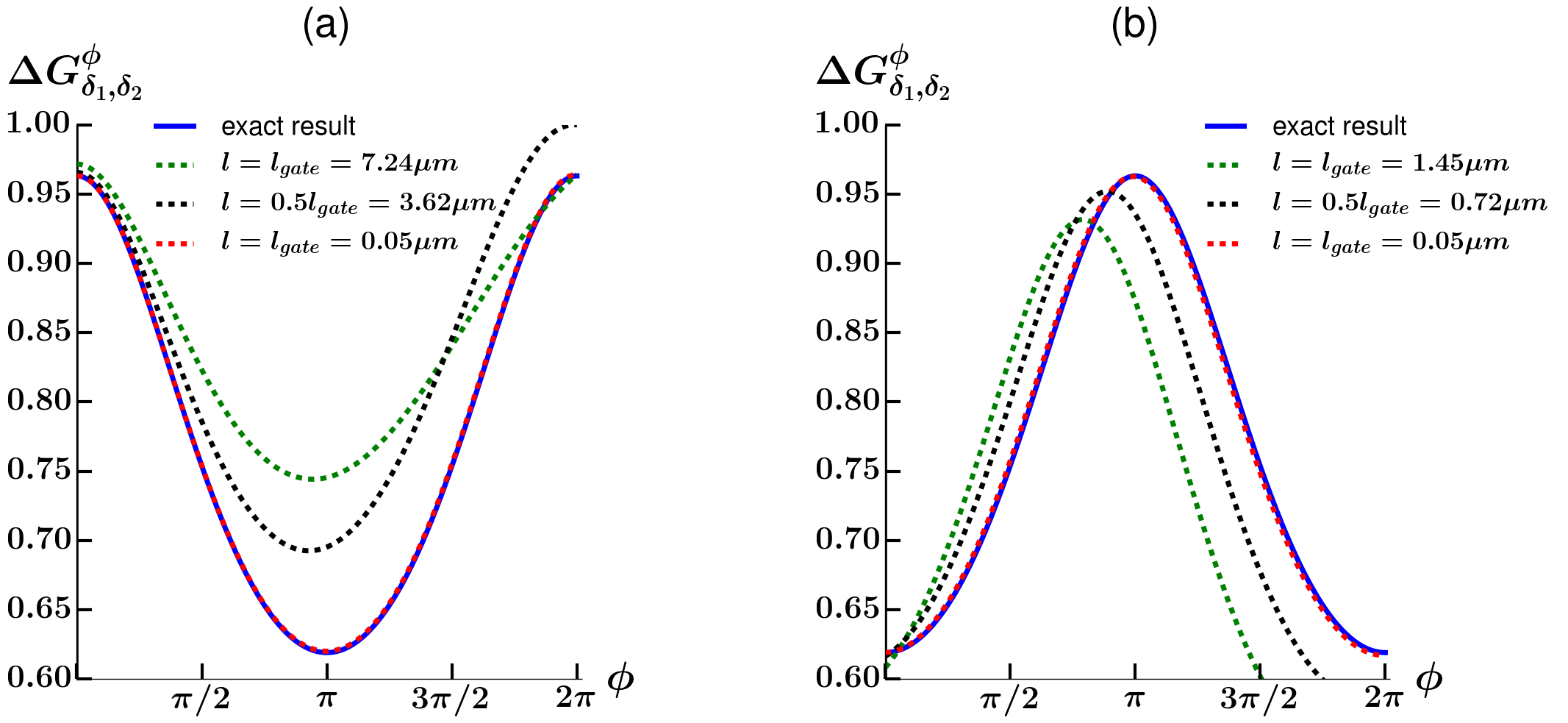}
\caption{Plot for Landauer conductance $G^{\phi}_{\delta_1,\delta_2}$ (calculated numerically) in units $e^2/h$ as a function of rotation of magnetic   field $\phi$ for different values of junction length $l$ in case of (a) $n$-$n$ junction and (b) $n$-$p$ junction. Here,we have used $V_{{{\cal G}}_1}=0.1mV$, $V_{{{\cal G}}_2}=0.15mV$, $B=50mT$, the effective Land\'e g-factor = 50. \underline{Note} : $l_{B}=2.5\mu m$.}
 \label{fig3}
\end{figure}
\\The typical value of $l_{gate}$ is $7.24\mu m$ in case of $n$-$n$ junction and is $1.45\mu m$ in case of $n$-$p$ junction respectively while, $l_{B}$ is $1.25\mu m$ for both the cases as the strength of the applied magnetic field remains same in both the $n$-$n$ and the $n$-$p$ junctions. In fig.(\ref{fig3}), we observe that a linear variation of gate voltage or the $\vec{B}$ which happens over length $l \sim 0.05 \mu m (<< l_{gate},l_{B})$ can be taken to be of step function form as far as conductance is concerned. Note that for this value of $l$, the results obtained numerically matches perfectly with the analytical results obtained in Eq.(\ref{T}). This length scale seems to be feasible within the present day experimental developments.
\end{document}